\documentclass[amstex,aps,showpacs]{revtex4}%
\usepackage{graphicx}
\usepackage{amsmath}
\usepackage{bm}%
\usepackage{amsfonts}%
\usepackage{amssymb}

\begin{document}

\title{Contribution of forbidden orbits in the photoabsorption spectra of
atoms and molecules in a magnetic field}
\author{A.\ Matzkin}
\affiliation{Laboratoire de Spectrom\'{e}trie physique, CNRS and
Universit\'{e} Joseph-Fourier Grenoble-I, BP 87, F-38402
Saint-Martin, France}
\author{P.\ A.\ Dando}
\author{T.\ S.\ Monteiro}
\affiliation{Department of Physics and Astronomy, University
College London, Gower Street, London WC1E 6BT, Great Britain.}

\begin{abstract}
In a previous work [Phys.\ Rev. A \textbf{66}, 0134XX (2002)] we noted a
partial disagreement between quantum $R$-matrix and semiclassical calculations
of photoabsorption spectra of molecules in a magnetic field. We show this
disagreement is due to a non-vanishing contribution of processes which are
forbidden according to the usual semiclassical formalism. Formulas to include
these processes are obtained by using a refined stationary phase
approximation. The resulting higher order in $\hbar$ contributions also
account for previously unexplained ``recurrences without closed-orbits''.
Quantum and semiclassical photoabsorption spectra for Rydberg atoms and
molecules in a magnetic field are calculated and compared to assess the
validity of the first-order forbidden orbit contributions.

\end{abstract}
\maketitle


\bigskip

\section{Introduction}

\label{intro}The photo-absorption spectrum of excited atoms or
molecules placed in a magnetic field displays complex structures.
Closed-orbit theory consists of a fully quantitative approach in
which the large-scale structures of the spectra are explained in
terms of classical trajectories closed at the nucleus, i.e.,
leaving and returning to the core.\ Each orbit produces on its
return an oscillation in the photo-absorption cross-section; the
Fourier-transform of the spectrum, known as the \emph{recurrence
spectrum\/} therefore exhibits sharp peaks at the period of the
orbits. First developed for the hydrogen atom \cite{du delos88},
closed-orbit theory was then extended to treat the case of
non-hydrogenic Rydberg atoms: the additional spectral modulations
appear as the result of successive quantum encounters of the
Rydberg electron with the core \cite{dando etal95}. The
wave-function follows the \emph{hydrogenic\/} classical orbits in
the region where the Coulomb and the external fields compete
(``outer region''), but near the core, where the external fields
are negligible (``inner region''), the wave-function is described
quantum-mechanically.

More recently, we proposed a closed-orbit theory treatment of
molecules in external fields \cite{m&m01,mdm02}: in addition to
elastic scattering, inelastic scattering gives rise to novel
spectral modulations. The relative importance of elastic and
inelastic scattering was shown to depend on the short-range
phase-shifts, the molecular quantum defects. In the inelastic
collision process, the molecular core undergoes a transition from
its ground state to an excited state, and the dynamical regime of
the Rydberg electron changes accordingly, say from a chaotic to a
near integrable classical regime. Good quantitative agreements
between quantum calculations and closed-orbit theory in the case
of an external magnetic field were obtained. However we noted in
Ref. \cite{mdm02} a discrepancy between the quantum and the
semiclassical results for certain peaks of the recurrence
spectrum. In general, disagreements between semiclassical and
quantum recurrence spectra are due to higher order (in $\hbar$)
effects such as bifurcations or ghost orbits, and specific
formulas to account for these effects have been developed
\cite{sadovskii delos96,main wunner97}. The discrepancies we
observed in \cite{mdm02} are due to another type of effect, namely
the manifestation of orbits which are forbidden to first order in
$\hbar;$ these orbits are forbidden either \emph{i)\/} because they
should not be excited according to the usual semiclassical
formalism or \emph{ii)\/} because they do not classically exist.

We investigate in this work the effects of such first-order
suppressed orbits. We will derive formulas for including their
contribution in the recurrence spectra by going beyond the usual
stationary phase approximation employed in the standard form of
closed-orbit theory.\ The formulas will be tested versus exact quantum
calculations for different model atomic and molecular systems in
an external magnetic field. The two orbits that will be dealt with
specifically are the orbit perpendicular to the field, which should
not be excited when it lies in the node of a wavefunction, and the
orbit parallel to the field, which does not exist classically when
the electron's angular momentum projection on the field axis
$L_{z}$ is non-vanishing (since the Hamiltonian remains divergent
on the $z$ axis even after regularization).
The contribution of orbits lying in the
node of a wave-function was first observed by Shaw et al.
\cite{shaw etal95} when comparing quantum and semiclassical
calculations for the diamagnetic hydrogen atom in the
near-integrable regime (scaled energies $\epsilon\simeq-0.7$).
They obtained a formula for including the first-order forbidden
contribution of the perpendicular orbit, which appeared as a small
feature in the recurrence spectra.\ In non-hydrogenic systems, we
can expect the effects associated with ``forbidden'' orbits to be
far more important than in hydrogen given that core-scattering
mixes the contributions of different orbits. Moreover, although it
may have been expected that at higher scaled energy the
contribution of on-node orbits would become insignificant (as the
classical amplitudes decrease), we will see that their inclusion
is necessary to account for the correct amplitude in the
modulations produced by orbits which have bifurcated from them.
The existence of recurrences produced by ``non-existing'' orbits
on the field axis was reported for Rydberg atoms in an electric
field by Robicheaux and Shaw \cite{robicheaux shaw98}, who
developed a heuristic formula which yielded a poor agreement
between semiclassical and quantum calculations. We will derive a
formula for appropriately taking into account such classically
forbidden orbits and compare it to quantum calculations in the
case of an external magnetic field.\ In passing we will also show
that the contribution of the parallel orbit when it is allowed
(i.e., for $L_{z}=0$) can be obtained by treating it as any other
orbit, provided a higher order refined stationary phase
integration is used (whereas the parallel orbit has always been
treated as a special case, following the original derivation given
by Gao and Delos \cite{gao delos92}).

The paper is organized as follows.\ We recall in Sec.\ II the
usual semiclassical formulas of closed-orbit theory (with
provision for multichannel core scattering).\ Sec.\ III details
the derivation of the contribution to the photoabsorption spectra
of the first-order forbidden orbits. We report in Sec.\ IV quantum
and semiclassical calculations for different values of the quantum
defects, scaled energies or magnetic field ranges, focusing on the
contribution of those forbidden orbits.\ We give our conclusions
in Sec.\ V.

\section{Overview of standard multichannel closed-orbit theory}

\label{COT}

Closed-orbit theory explains the dynamics underlying the
photoabsorption spectra of Rydberg atoms or molecules in external
fields in terms of closed orbits: following initial
photo-excitation, the wavefunction of the excited electron
propagates first in a region near the ionic core (``inner
region''), in which the external field can be neglected. Beyond
the inner region, the wave-function is propagated semiclassically
along classical trajectories. Some trajectories return to the
inner region, and the semiclassical wavefunction carried by those
trajectories is matched to an exact wavefunction in the inner
region given by a standard (field-free) multichannel quantum
defect theory (MQDT) expansion. The superposition of these
returning waves with the initially dipole-excited wavefunction
produces sinusoidal modulations in the photoabsoprtion spectrum,
which appear as isolated peaks in the Fourier-transformed
(``recurrence'') spectrum. Further modulations (i.e., peaks in the
recurrence spectrum) are caused by the core-scattering process; in
a multichannel problem, the electron can exchange energy and
angular momenta with the core, so after the collision the electron
wavefunction propagates outward again, and when it leaves the
inner region the wavefunction will follow once again classical
trajectories. If the collision was perfectly elastic, the electron
will follow one of the previously followed trajectories; on the
other hand, inelastic collisions will result in trajectories
pertaining to a different classical regime.

In Ref. \cite{mdm02} we described in detail photoabsorption from a
ground state diatomic molecule in a static magnetic field. After
photoexcitation, the molecular core could either be rotationally
excited ($N=2$) or non-excited $(N=0,$ where $N$ is the core
angular momentum).\ The molecular core then plays the role of an
effective 2-level scatterer which combines classical trajectories
belonging to 2 different dynamical regimes (typically, chaotic and
near-integrable regimes), thereby producing additional modulations
in the photoabsorption spectrum.\ The way these combinations occur
depends both on the classical characteristics (amplitude $A_{k}$
and action $S_{k}$ of the $k$th trajectory), and on the properties
of the scatterer (which is given by the scattering transition
matrix, $T$). Scaled energy spectroscopy consists of simultaneously
varying the magnetic field strength $\gamma$ and the laser
excitation frequency so as to keep $\epsilon=E\gamma^{-2/3}$
constant, where $E$ is the energy of the Rydberg electron;
$\epsilon$ is the scaled energy, which depends on the core state
$j$ through the energy partition between the core and the outer
electron. Although scaling for molecules is only approximate, we
have seen in \cite{mdm02} how a molecular system can be scaled
conveniently; $\hbar_{\textrm{eff}}$ will stand for
$\gamma^{1/3},$ since the field strength plays the r\^{o}le of the
Planck constant \cite{friedrich wintgen89}. The absorption rate in
the one core-scatter approximation is then
given by [see Eq. (3.30) in \cite{mdm02}]%
\begin{align}
\mathcal{F}(\hbar_{\text{eff}})=2^{19/4}\pi^{3/2}\sum_{j}\sum_{\alpha}%
\sum_{\alpha^{\prime}}\operatorname{Im}\left\{  \left\langle
\alpha\right|
\left.  j\right\rangle \rm{C}_{\alpha}\rm{C}_{\alpha^{\prime}%
}e^{i\pi\left(  \mu_{\alpha}+\mu_{\alpha^{\prime}}\right)
}\right.
\nonumber\\
\left.  \left[  \left\langle j\right|  \left.
\alpha^{\prime}\right\rangle
\sum_{k}\tilde{\mathcal{R}}_{k}^{j}(\epsilon_{j})+\hbar_{\text{eff}}%
^{1/2}2^{11/4}\pi^{3/2}\sum_{j^{\prime}}\left\langle
j^{\prime}\right| \left.  \alpha^{\prime}\right\rangle
T_{jj^{\prime}}\sum_{k}\tilde
{\mathcal{R}}_{k}^{j^{\prime}}(\epsilon_{j^{\prime}})\sum_{q}\tilde
{\mathcal{R}}_{q}^{j}(\epsilon_{j})\right]  \right\}  .\label{10}%
\end{align}
The notation has been completely detailed in Sec.\ III of
\cite{mdm02}, but in short: $j$ (and $j^{\prime}$) is a compound
index accounting for the core-electron couplings when the electron
dynamics has uncoupled from the core. If we assume H$_{2}$ as the
prototype molecule, the initial state has the quantum numbers
$J=0,$ $l=0$ ($J$ is the total angular momentum, $l$ the orbital
momentum of the outer electron), the sum over $j$ runs over the
core states $N=0,$ $M_{N}=M$ and $N=2,$ $M_{N}=M-1,M,M+1$ where
$M_{N}$ is the projection of $N$ on the field axis; $M,$ the
projection of the total angular momentum on the field axis, is the
only quantum number conserved throughout the entire physical
process. $\alpha$ gives the set of quantum numbers in the
molecular frame, when the electron is coupled to the molecular
axis; the sum runs here on $\Lambda=0$ ($\Sigma$ state) and
$\Lambda=1$ ($\Pi$ state); $\mu_{\Sigma}$ and $\mu_{\Pi}$ are the
corresponding short-range molecular quantum defects.
$\rm{C}_{\alpha}$ is a coefficient giving the relative strength of
the electronic dipole transition amplitudes; from the united
dipole approximation, known to be valid for H$_{2},$ we have $\rm{C}%
_{\Sigma}=1$ and $\rm{C}_{\Pi}=\sqrt{2}.$ $\left\langle j\right|
\left. \alpha\right\rangle $ are the transformation coefficients
between the molecular and the uncoupled frames. The elements of
the scattering matrix $T_{jj^{\prime}}$ depend solely upon the
quantum defects and the $\left\langle j\right|  \left.
\alpha\right\rangle $ elements. The quantities
$\tilde{\mathcal{R}}_{k}^{j}(\epsilon_{j})$ are the only ones that
depend on the classical properties of the Rydberg electron
trajectories. We have, for the trajectory $k$ associated with the
core in state $j$ [Eqs. (3.18) and (D1) of \cite{mdm02}]:%
\begin{align}
\tilde{\mathcal{R}}_{k}^{j}(\epsilon_{j})=\left|
\sin\theta_{ik}\sin \theta_{fk}\right|
^{1/2}\sum_{l_{j}%
l_{j^{\prime}}}(-1)^{l_{j}+l_{j^{\prime}}}%
Y_{l_{j^{\prime}}m_{j^{\prime}}}(\theta_{ik})Y_{l_{j}m_{j}}^{\ast}(\theta
_{fk})\nonumber\\
\tilde{r}_{f}^{-1/4}A_{k}^{N_{j}m_{j}}(r_{f},\theta_{fk})e^{i\left(  2\pi\tilde{S}_{k}^{N_{j}%
m_{j}}/\hbar_{\text{eff}}-\omega_{k}^{N_{j}m_{j}}\pi/2-3\pi/4\right)
},\label{11}%
\end{align}
where $l_j,l_{j^{\prime}}\geqslant\left| m_{j}\right|$ and
$m_{j^{\prime}}=m_{j}$. $\theta_{ik}$ and $\theta_{fk}$ are the
initial and final angles of the $k$th trajectory relative to the
magnetic field direction, which is taken to be along the $z$ axis.
$A_{k}^{N_{j}m_{j}}$ and $\tilde{S}_{k}^{N_{j}m_{j}}$ are the
scaled classical amplitude and action, evaluated at the
corresponding scaled energy $\epsilon_{j}$;
$\omega_{k}^{N_{j}m_{j}}$ is the associated Maslov index. $Y_{l_{j}m_{j}}%
(\theta_{ik})$ will be used throughout as a short-hand notation
for $Y_{l_{j}m_{j}}(\theta_{ik},0),$ since the conserved axial
symmetry has been separated from the 2-dimensional semiclassical
problem; we have accordingly used a different notation for the
quantized value of the Rydberg electron's angular momentum
projection $m$ and its classical counterpart $L_{z}$ appearing in
the two-dimensional diamagnetic Hamiltonian. Equation (\ref{11}) must
be modified for orbits lying along the magnetic field axis
($\theta _{ik}=\theta_{fk}=0$) as detailed below. Note that Eq.
(\ref{10}) is also valid for ground state hydrogen photoexcited to
odd-parity states (by setting $\mu_{\Sigma}$ and $\mu_{\Pi}$ to
zero; then the $T$ matrix vanishes) as well as for non-hydrogenic
Rydberg
atoms with a single quantum defect $\mu_{l=1}$ (by setting $\mu_{\Sigma}%
=\mu_{\Pi}=\mu_{l=1}$; the $T$ matrix is then diagonal).

Eqs. (\ref{10}) and (\ref{11}) are obtained by matching the semiclassical
wavefunction $\psi_{SC}^{N_{j}m_{j}}$ associated with the core in state
$\left|  N_{j}m_{j}\right\rangle $ which returns to the core region to a MQDT
expansion on a boundary circle $(r_{f},\theta_{f})$. The semiclassical
wavefunction reads%
\begin{equation}
\psi_{SC}^{N_{j}m_{j}}(r_{f},\theta_{f})=\sum_{k}\psi_{\mathrm{out}}%
^{N_{j}m_{j}}(r_{i},\theta_{ik})\left|  \frac{r_{i}^{2}\sin\theta_{ik}}%
{r_{f}^{2}\sin\theta_{f}}\right|  ^{1/2}A_{k}^{N_{j}m_{j}}(r_{f},\theta
_{f})\exp i\left(  S_{k}^{N_{j}m_{j}}(r_{f},\theta_{f})-\omega_{k}^{N_{j}%
m_{j}}\pi/2\right)  ,\label{15}%
\end{equation}
where $\psi_{\mathrm{out}}^{N_{j}m_{j}}(r_{i},\theta_{ik})$ represents the
initially outgoing waves, which were propagated semiclassically beyond the
boundary $(r_{i},\theta_{i}).$ We shall write $\psi_{\mathrm{out}}^{N_{j}%
m_{j}}(r_{i},\theta_{ik})$ as
\begin{equation}
\psi_{\mathrm{out}}^{N_{j}m_{j}}(r_{i},\theta_{ik})=\sum_{l_j\geqslant\left|
m_{j}\right|  }%
\mathcal{Q}_{l_{j}}Y_{l_{j}m_{j}}(\theta_{ik}),\label{16}%
\end{equation}
with
\begin{equation}
\mathcal{Q}_{l_{j}}=-i\pi^{1/2}2^{3/4}r_{i}^{-3/4}(-1)^{l_{j}}%
e^{i\left(  \sqrt{8r_{i}}-3\pi/4\right)  }%
\sum_{\alpha}e^{i\pi\mu_{\alpha}}\left\langle N_{j}l_{j}m_{j}\right|  \left.
\alpha\right\rangle D_{\alpha},
\end{equation}
where the $D_{\alpha}$ are the dipole transition amplitudes in the molecular
frame. The MQDT expansion reads in the uncoupled basis%
\begin{equation}
\psi_{\mathrm{qdt}}(r_{f})=\sum_{j}\left|  j\right\rangle \sum_{j^{\prime}%
}c_{j^{\prime}}\left[  \delta_{jj^{\prime}}f_{l_{j}}(r_{f})+T_{jj^{\prime}%
}g_{l_{j}}^{+}(r_{f})\right]  ,\label{17}%
\end{equation}
where $f$ and $g^{+}$ are Coulomb functions ($f$ is regular at the origin,
$g^{+}$ is an outgoing wave).\ The expansion coefficients $c_{j^{\prime}}$ are
obtained by matching Eqs. (\ref{15}) and (\ref{17}) on the boundary. The
matching condition reads%
\begin{equation}
c_{j}g_{l_{j}}^{-}(r_{f})/2i=-2\pi\int_{0}^{\pi}d\theta_{f}\,\sin\theta
_{f}Y_{l_{j}m_{j}}^{\ast}(\theta_{f})\psi^{N_{j}m_{j}}(r_{f},\theta
_{f});\label{19}%
\end{equation}
the integral is performed for each trajectory $k$ in the
\emph{stationary phase approximation\/}, since the phase is
stationary along the final angle of the trajectory $\theta_{fk}$
\cite{hupper etal96}. The value of the coefficients $c_{j}$ are
then inserted in the expression giving the dipole transition
amplitudes, of the form $\left\langle \psi_{0}\right|
\mathbf{D}\left| \psi_{\mathrm{qdt}}\right\rangle $ where $\left|
\psi_{0}\right\rangle $ is the initial state prior to
photoabsorption.\ The formulas for the oscillator strength and the
absorption rate are then obtained. Obviously when Eqs. (\ref{15})
or (\ref{19}) vanish, e.g., $Y_{l_{j}m_{j}}(\theta_{ik})=0$ or
$Y_{l_{j}m_{j}}^{\ast}(\theta_{f})=0$, then $\tilde{\mathcal{R}}_{k}%
^{j}(\epsilon_{j})$ vanishes, and these orbits should not produce modulations
in the oscillator strength.

\section{Contribution of first-order forbidden orbits}

\subsection{General remarks}

Although Eqs. (\ref{10}) and (\ref{11}) predict that if $\tilde{\mathcal{R}%
}_{k}^{j}(\epsilon_{j})$ vanishes, the orbit $k$ should not
contribute to the recurrence spectrum, we had observed in
\cite{mdm02} a mismatch between semiclassical and quantum
classical calculations for molecules in fields in the amplitude of
certain peaks in the recurrence spectrum. This mismatch was
interpreted as arising from the interference of the orbit
perpendicular to the field (which lies on the node of a spherical
harmonic when $M=0$ and is thus semiclassically forbidden) with
the $R_{2}^{1}$ ``pac-man'' orbit. As stated, the recurrences
associated with classical orbits lying in the node of a
wavefunction were first observed by Shaw et al. \cite{shaw etal95}
when comparing quantum and semiclassical calculations for the
diamagnetic hydrogen atom at low scaled energies
($\epsilon\simeq-0.7$). They obtained a formula for the
contribution of the perpendicular orbit by matching the returning
semiclassical wave to an ``ansatz'' (a rotated first order Bessel
function). In this section we shall derive simply the contribution
to the oscillator strength of this type of first-order suppressed
orbit by employing the same framework introduced in Sec.\ II,
without introducing additional assumptions; only the stationary
phase integration needs to be performed differently.\ We will also
derive a formula to account for the peaks in the $m=1$ recurrence
spectra appearing at the scaled action of the $m=0$ parallel
orbit, since the parallel orbit does not exist classically when
$L_{z}\neq0$ and there is therefore no corresponding
$\tilde{\mathcal{R}}_{k}^{j}(\epsilon_{j})$ factor. Numerical
results and examples will be given in Sec. IV.

\subsection{Contribution of on-node suppressed orbits\label{sec3b}}

The rationale for including the contribution of orbits lying on the node of a
wavefunction was already given in \cite{shaw etal95}: strictly speaking, an
orbit $k$ closed at the core with initial and returning angles $\theta_{ik}$
and $\theta_{fk}$ is not isolated, but has neighboring orbits which are not
closed at the origin.\ We assume a neighboring orbit returns with an angle
$\bar{\theta}_{f}$ and envisage the initial angle $\bar{\theta}_{i}$ of this
orbit to be a function of $\bar{\theta}_{f},$ i.e., $\bar{\theta}_{i}%
=\theta(\bar{\theta}_{f})$. To first order in $\bar{\theta}_{f}-\theta_{fk}$
we have%
\begin{align}
Y_{l_{j}m_{j}}(\bar{\theta}_{i}) &  =Y_{l_{j}m_{j}}(\theta_{ik}%
)+\frac{\partial Y_{l_{j}m_{j}}(\theta_{ik})}{\partial\theta_{ik}}\left.
\frac{\partial\theta_{ik}}{\partial\bar{\theta}_{f}}\right|  _{\theta_{fk}%
}\left(  \bar{\theta}_{f}-\theta_{fk}\right) \label{20}\\
Y_{l_{j}m_{j}}^{\ast}(\bar{\theta}_{f}) &  =Y_{l_{j}m_{j}}^{\ast}(\theta
_{fk})+\left.  \frac{\partial Y_{l_{j}m_{j}}^{\ast}(\bar{\theta}_{f}%
)}{\partial\bar{\theta}_{f}}\right|  _{\theta_{fk}}\left(  \bar{\theta}%
_{f}-\theta_{fk}\right) \label{21}%
\end{align}
In the usual case, the contribution of the neighboring orbits with initial and
final angles $(\bar{\theta}_{i},\bar{\theta}_{f})$ is negligible when compared
to the central orbit with angles $(\theta_{ik},\theta_{fk})$. However, when
the central orbit lies on a node of a spherical harmonic, the semiclassical
wave-function can only be carried by the neighboring orbits, and this
contribution can be significant provided the classical density of trajectories
is sufficiently large on return to the core.

It turns out that, to first order in
$\bar{\theta}_{f}-\theta_{fk}$, such a contribution comes into
play if we
have both $Y_{l_{j}m_{j}}(\theta_{ik})=0$ and $Y_{l_{j}m_{j}}^{\ast}%
(\theta_{fk})=0$. If  $k$ is such an orbit, its contribution to
the outgoing wave $\psi_{\mathrm{out}}^{N_{j}m_{j}}$ [Eq.
(\ref{16})] vanishes. The contribution of the neighboring orbits
are taken into account by inserting Eq. (\ref{20}) in Eq.
(\ref{16}) and Eq. (\ref{21}) in Eq. (\ref{19}); the right
hand-side of Eq. (\ref{19}) then takes the
form:%
\begin{align}
-2\pi\int_{0}^{\pi}d\bar{\theta}_{f}\left|  \sin\bar{\theta}_{f}\sin
\theta_{ik}\right|  ^{1/2}\sum_{l_{j^{\prime}}}\mathcal{Q}_{l_{j^{\prime}}%
}\frac{\partial Y_{l_{j^{\prime}}m_{j}}(\theta_{ik})}{\partial\theta_{ik}%
}\left.  \frac{\partial\theta_{ik}}{\partial\bar{\theta}_{f}}\right|
_{\theta_{fk}}\left.  \frac{\partial Y_{l_{j}m_{j}}^{\ast}(\bar{\theta}_{f}%
)}{\partial\bar{\theta}_{f}}\right|  _{\theta_{fk}} & \nonumber\\
A_{k}^{N_{j}m_{j}}(r_{f},\bar{\theta}_{f})\left[  \left(  \bar{\theta}%
_{f}-\theta_{fk}\right)  ^{2}\exp i\left(  S_{k}^{N_{j}m_{j}}(r_{f}%
,\bar{\theta}_{f})-\omega_{k}^{N_{j}m_{j}}\pi/2\right)  \right]  . &
\label{25}%
\end{align}
Following H\"{u}pper et al. \cite{hupper etal96}, we express the action on the
boundary $(r_{f},\bar{\theta}_{f})$ in terms of the action of the orbit closed
at the origin, $S_{k}^{N_{j}m_{j}}(r_{f},\bar{\theta}_{f})\simeq
S_{k\mathrm{(closed)}}^{N_{j}m_{j}}+\sqrt{r_{f}/8}\left(  \bar{\theta}%
_{f}-\theta_{fk}\right)  ^{2}$. The integral can now be performed; a
straightforward stationary phase integration would lead to zero, since the
integrand vanishes at the point of stationary phase $\bar{\theta}_{f}%
=\theta_{fk}$.\ However, we can assume the integrand to vary slowly around the
angle of stationary phase, and integrate exactly the term between square
brackets (see Appendix A). In the semiclassical limit, Eq. (\ref{a6}) is
appropriate.\ Eq. (\ref{25}) then becomes%
\begin{align}
&  \hbar e^{i\pi/2}2^{1/2}\left[  r_{f}^{-1/4}A_{k}^{N_{j}m_{j}}(r_{f}%
,\theta_{fk})\right]  ^{2}\mathrm{sgn}(\frac{\partial\theta_{ik}}{\partial
{\theta}_{fk}})\nonumber\\
&  \left\{  -2^{3/4}\pi\left(  2\pi\hbar\right)  ^{1/2}e^{i\pi/4}%
\sum_{l_{j^{\prime}}}\mathcal{Q}_{l_{j^{\prime}}}\frac{\partial
Y_{l_{j^{\prime}}m_{j}}(\theta_{ik})}{\partial\theta_{ik}}\left.
\frac{\partial Y_{l_{j}m_{j}}^{\ast}(\bar{\theta}_{f})}{\partial\bar{\theta
}_{f}}\right|  _{\theta_{fk}}\right. \nonumber\\
&  \left.  \left|  \sin\theta_{ik}\sin\theta_{fk}\right|  ^{1/2}r_{f}%
^{-1/4}A_{k}^{N_{j}m_{j}}(r_{f},\theta_{fk})\exp\left[  i\left(
S_{k\mathrm{(closed)}}^{N_{j}m_{j}}-2\sqrt{8r_{f}}-\omega_{k}^{N_{j}m_{j}}%
\pi/2\right)  \right]  \right\}  ,\label{27}%
\end{align}
where we have used $A_{k}^{N_{j}m_{j}}(r_{f},\theta_{fk})=\left|
\frac{\partial\theta_{ik}}{\partial{\theta}_{fk}}\right| ^{1/2}$.
For clarity we have singled out the factor specific to on-node
orbits (in front of the curly brackets) relative to the expression
valid for ``typical'' allowed orbits (inside the curly brackets).
In particular it can be seen on-node orbits are suppressed by a
factor $\hbar$ relative to typical orbits.

The relevant scaled factor $\tilde{\mathcal{R}}_{k\mathrm{-node}}^{j}%
(\epsilon_{j})$ giving the contribution of an orbit lying on the node of a
wavefunction in the absorption rate is thus%
\begin{align}
\tilde{\mathcal{R}}_{k\mathrm{-node}}^{j}(\epsilon_{j}) &  =\hbar_{\textrm{eff}}%
2^{1/2}\mathrm{sgn}(\frac{\partial\theta_{ik}}{\partial{\theta}_{fk}}%
)\left|  \sin\theta_{ik}\sin\theta_{fk}\right|  ^{1/2}\sum_{l_{j}%
l_{j^{\prime}}}(-1)^{l_{j}+l_{j^{\prime}}}\frac{\partial Y_{l_{j^{\prime}%
}m_{j}}(\theta_{ik})}{\partial\theta_{ik}}\left.  \frac{\partial Y_{l_{j}%
m_{j}}^{\ast}(\bar{\theta}_{f})}{\partial\bar{\theta}_{f}}\right|
_{\theta_{fk}}\nonumber\\
&  \left[  \tilde{r}_{f}^{-1/4}A_{k}^{N_{j}m_{j}}(r_{f},\theta_{fk})\right]
^{3}\exp i\left(  2\pi\tilde{S}_{k}^{N_{j}m_{j}}/\hbar_{\textrm{eff}}-\omega_{k}%
^{N_{j}m_{j}}\pi/2-\pi/4\right)  .\label{29}%
\end{align}
This formula holds for non-vanishing angles. Specializing to
our atomic and molecular model described above, we have $l=1,$ so
this formula only applies to the orbit perpendicular to the field
($\theta_{ik}=\theta_{fk}=\pi/2$) for the $m=0$ manifolds (e.g.,
for a molecule, when $M=0,$ for an outer electron associated with
core states having a projection $M_{N}=0$). Note that in the
absence of core-effects, Eq. (\ref{25}) becomes strictly
equivalent to the correction obtained in \cite{shaw etal95} for
the hydrogen atom.

\subsection{Contribution of the ``classically non-existing'' parallel orbit}

\subsubsection{Contribution of the parallel orbit when $L_{z}=0$}

We first recall that the orbit parallel to the field
($\theta_{ik}=\theta_{fk}=0)$ classically exists if $L_{z}=0$.
Even then, this orbit is treated as a special case, because the
formulas valid for the other orbits, Eqs. (\ref{10}) and
(\ref{11}), need to be modified. This modification was originally
obtained by matching the semiclassical returning wave to a
particular Bessel function on the $z$ axis \cite{gao delos92}. We
show here that the reason this orbit is ``special'' is that the
standard stationary phase approximation vanishes. Indeed, setting
$\left|  \sin\theta_{ik}/\sin\theta_{fk}\right|
^{1/2}\rightarrow\left|
\partial\theta_{ik}/\partial\theta_{fk}\right| ^{1/2}$ in the
outgoing wave (\ref{15}), the expression to be integrated
arising from the matching condition Eq. (\ref{19}) is%
\begin{equation}
\int_{0}^{\pi}d\bar{\theta}_{f}\sin\bar{\theta}_{f}\exp(i\sqrt{r_{f}/8}\bar
{\theta}_{f}^{2}/\hbar),
\end{equation}
which is zero in the standard stationary phase approximation.
However, an approximate closed form may be obtained (see Appendix
B). To first order
in $\hbar$, we have%
\begin{equation}
\int_{0}^{\pi }d\bar{\theta}_{f}\sin \bar{\theta}_{f}\exp (i\sqrt{r_{f}/8}\bar{%
\theta}_{f}^{2}/\hbar )\approx \hbar e^{i\pi /2}\sqrt{2/r}.
\end{equation}%
This result is reduced by a factor
$h^{1/2}e^{i\pi/4}2^{-1/4}\pi^{-1/2}$ relative to the standard
stationary phase integration for non-zero degree orbits, which is
exactly the result obtained in \cite{gao delos92}. The parallel
orbit thus appears as a first-order suppressed orbit, which is
apparent from its $\hbar$-dependence.

\begin{figure}[tb]
\includegraphics[height=2.in,width=2.5in]{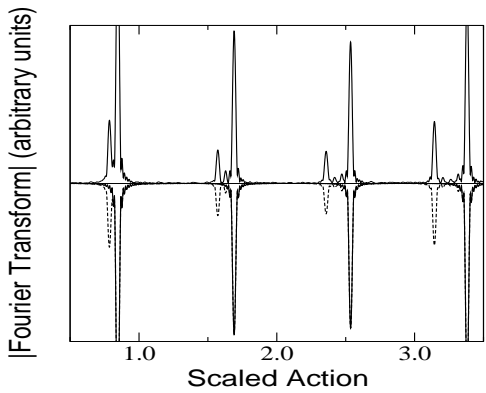}
\caption[]{Recurrence spectrum (Fourier transform of the
photoabsorption spectrum) for a non-hydrogenic atom with
$\mu_{l=1}=0.5,$ $M=0,$ at $\epsilon=-0.7$, in the range $\gamma
^{-1/3}=[60,120]$. Top: quantum calculations.\ Bottom: standard
semiclassical calculations (solid line), semiclassical calculation
including the higher-order contribution from the on-node orbit
(broken line). \label{fig.1}}
\end{figure}

\subsubsection{Contribution of the parallel orbit when $L_{z}\neq0$}

When $L_{z}$ is non-vanishing the diamagnetic Hamiltonian contains
the repulsive term proportional to $\tilde{L}_{z}^{2}/\tilde{\rho}^{2}$,
where $\tilde {L}_{z}=\gamma^{1/3}L_{z}$ is the scaled angular
momentum and $\tilde{\rho }=\gamma^{2/3}\rho$ the scaled distance
from the $z$ axis \cite{friedrich wintgen89}. Even though in the
semiclassical limit $\tilde{L}_{z}$ is small (since
$h_{\mathrm{\textrm{eff}}}\equiv\gamma^{1/3}\rightarrow0$), the
centrifugal term is infinite on the $z$ axis, and the parallel
orbit no longer exists \cite{schweizer etal93}. However, we may
expect orbits neighboring the $z$ axis and not closed at the
nucleus to contribute to the oscillator strength, in the same
manner as for the on-node orbit (orbits near the $z$ axis for
$L_{z}\neq0$ and their structural stability as $L_{z}\rightarrow0$
where actually investigated in \cite{schweizer etal93}). Starting
from Eqs. (\ref{20}) and (\ref{21}) and setting $\left|
\sin\theta_{ik}/\sin\theta _{fk}\right|  ^{1/2}\rightarrow\left|
\partial\theta_{ik}/\partial\theta _{fk}\right|  ^{1/2}$ in Eq.
(\ref{15}) as in Sec. \ref{sec3b} leads, after matching the
semiclassical returning wave to the MQDT expansion, to the
integral%
\begin{equation}
\int_{0}^{\pi}d\bar{\theta}_{f}\sin\bar{\theta}_{f}\bar{\theta}_{f}^{2}\exp
(i\sqrt{r_{f}/8}\bar{\theta}_{f}^{2}/\hbar).
\end{equation}
To lowest order in $\hbar ,$ we have (see Appendix B)%
\begin{equation}
\int_{0}^{\pi }d\bar{\theta}_{f}\sin \bar{\theta}_{f}\bar{\theta}_{f}^{2}\exp (i%
\sqrt{r_{f}/8}\bar{\theta}_{f}^{2}/\hbar )\approx \frac{-4\hbar
^{2}}{r}.
\end{equation}%
The resulting scaled contribution to the oscillator strength is given by%
\begin{align}
&  \tilde{\mathcal{R}}_{k\mathrm{-forb\,0}}^{j}(\epsilon_{j})=\hbar
_{\textrm{eff}}^{3/2}2^{5/4}\pi^{-1/2}\mathrm{sgn}(\frac{\partial\theta_{ik}}%
{\partial{\theta}_{fk}})\sum_{l_{j}l_{j^{\prime}}}(-1)^{l_{j}%
+l_{j^{\prime}}}\nonumber\\
&  \frac{\partial Y_{l_{j^{\prime}}m_{j}}(\theta_{ik})}{\partial\theta_{ik}%
}\left.  \frac{\partial Y_{l_{j}m_{j}}^{\ast}(\bar{\theta}_{f})}{\partial
\bar{\theta}_{f}}\right|  _{\theta_{fk}}\left[  \tilde{r}_{f}^{-1/2}%
A_{k}^{N_{j}m_{j}}(r_{f},\theta_{fk}=0)\right]  ^{2}\nonumber\\
&  \exp i\left(  2\pi\tilde{S}_{k}^{N_{j}m_{j}}/\hbar_{\textrm{eff}}-\omega_{k}%
^{N_{j}m_{j}}\pi/2\right)  .\label{31}%
\end{align}
Within our molecular model, this correction applies when $\left|  m\right|
=1$; this is of course the case when $M=1$, but even for $M=0,$ the outer
electron may be associated with core states having a projection $\left|
M_{N}\right|  =1$, i.e., $N=2$ $M_{N}=1,-1$. Note that we have followed the by
now standard notation whereby the zero-degree orbit amplitude is set as
$A_{k}^{N_{j}m_{j}}(r_{f},\theta_{fk}=0)=\left|  \partial\theta_{ik}%
/\partial\theta_{fk}\right|  $ (although stricto sensu this is the square of
the genuine two-dimensional semiclassical amplitude), so that now $\tilde
{r}_{f}^{-1/2}A_{k}^{N_{j}m_{j}}(r_{f},\theta_{fk}=0)$ is independent of the
boundary radius $r_{f}$.

\subsection{$\hbar$ dependence}

Unsurprisingly, the contribution of the forbidden orbits in the
recurrence spectra have a different $\hbar$ dependence. The
on-node orbit is suppressed by a factor $\hbar$ relative to a
typical primitive orbit; the parallel orbit is suppressed by a
factor $\hbar^{1/2}$ relative to a typical orbit, and the
forbidden parallel by a factor $\hbar$ relative to the classically
allowed parallel orbit and $\hbar^{3/2}$ relative to a typical
orbit. This is to be contrasted with the core-scattered
(``diffractive'') orbits: each encounter with the core brings in
for a typical orbit a factor $\hbar^{1/2}$. Thus single
core-scattering is expected to dominate the photoabsorption
spectrum in the semiclassical regime; however, the $\hbar$
dependence is balanced by the amplitude factors, explaining why
for individual orbits the forbidden contribution may be strong, as
will be seen below. It may also be noted that the combination of
orbits having different individual $\hbar$ dependence through
core-scattering [last term in Eq. (\ref{10})] will give rise to
peaks in the recurrence spectra with a dependence of the form
$\hbar^{\nu/2}$, where $\nu$ is an integer depending on the type
of primitive orbits connected by the core-scattering process. In
particular, core-scattering between two forbidden parallel orbits
is expected to be highly suppressed in the semiclassical limit.

\begin{figure}[tb]
\includegraphics[height=2.15in,width=3in]{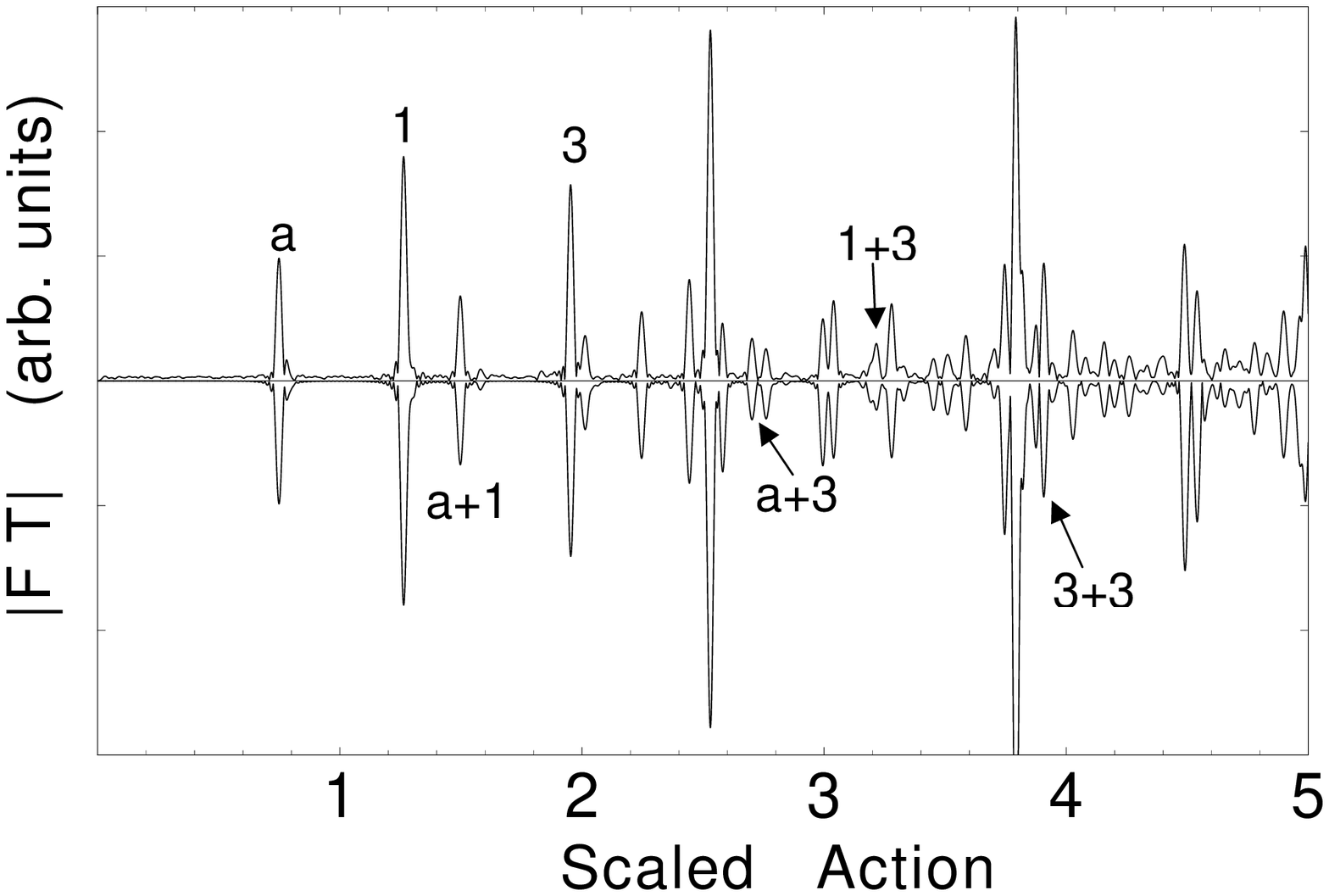}
\caption[]{General view of the recurrence spectrum for a molecule
with the set of quantum defects $\mu_{\Sigma}=-0.3,$
$\mu_{\Pi}=0.1,$ $M=0$, at $\epsilon_{N=0}=-0.3$ and
$\epsilon_{N=2}=-0.8$ in the range $\gamma ^{-1/3}=[60,120]$. The
semiclassical result (bottom) has been calculated in the one
core-scatter approximation, but the first-order suppressed
contributions have been included. \label{fig.2}}
\end{figure}
\section{Results}

We compare below quantum and semiclassical calculations to assess the
importance of the forbidden orbits in the recurrence spectra of atoms and
molecules. The numerical examples given in this section correspond to
non-hydrogenic atoms and different molecules obtained by choosing different
sets of quantum defects, within the framework of the model described in Sec.
\ref{COT}.

Fig.\ 1 displays the recurrence spectrum of a non-hydrogenic atom
with $\mu_{l=1}=0.5,$ $M=0,$ at $\epsilon=-0.7$, in the range
$\gamma ^{-1/3}=[60,120]$.\ The top figure gives the quantum
calculation, whereas the solid line in the bottom part of the plot
results from the standard semiclassical treatment; this solid line
only accounts for less than half of the peaks in the recurrence
spectrum. The missing peaks relative to the quantum results arise
from the orbit $R_1$ perpendicular to the field (and its $n$th
repetition $R_{n}$) --- which lies on the node of the wavefunction
and is thus not excited according to the standard treatment --- as
well as from the combinations produced by core-scattering between $R_{n}$
and the parallel orbit and between the on-node orbits. The broken
line includes the contribution of the on-node orbit [Eq.
(\ref{29})] in the semiclassical calculation.

\begin{figure}[tb]
\includegraphics[height=3.in,width=1.85in]{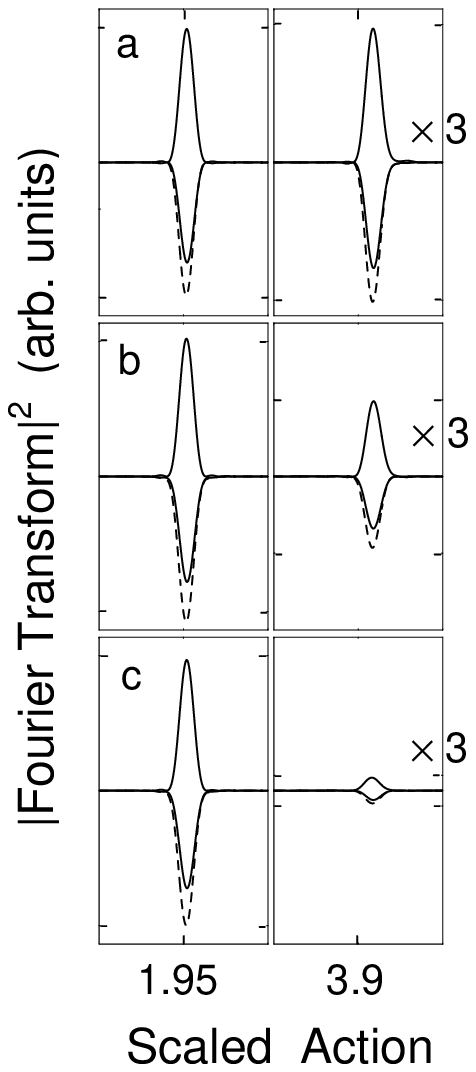}
\caption[]{Recurrence spectra for non-hydrogenic atoms with
$\mu_{l=1}=0.5$ [a]$,$ $\mu_{l=1}=0.25$ [b] and $\mu_{l=1}=0.1$
[c]. The left panel represents the peak labelled $3$ in Fig.\ 2,
the right panel shows the combination peak $1+3$ (the amplitude of
the $1+3$ peak has been multiplied by 3 relative to the amplitude
of the peak on the left panel). For each peak, the quantum result
(top) is plotted versus semiclassical calculations (upside-down)
without (solid line) and with (broken line) the higher order
contributions. \label{fig.3}}
\end{figure}
At higher scaled energies, the contribution of the forbidden
$R_{n}$ orbit is visible through the mismatch observed in
\cite{mdm02} between the height of the peaks in the quantum and
semiclassical recurrence spectra. Fig.\ 2 displays a global view
of the recurrence spectrum for a molecule with the set of quantum
defects $\mu_{\Sigma}=-0.3,$ $\mu_{\Pi}=0.1,$ $M=0$, at
$\epsilon_{N=0}=-0.3$ and $\epsilon_{N=2}=-0.8$ in the range
$\gamma ^{-1/3}=[60,120]$. These quantum defects yield a balanced
contribution of the different type of orbits: the primitive
geometric orbits (that is the orbits that appear in the recurrence
spectrum of the hydrogen atom), the elastic scattered diffractive
orbits (that appear in the recurrence spectra of non-hydrogenic
atoms and in molecules) and the inelastic scattered diffractive
orbits (that solely appear in molecular systems). In Figs.\ 3--5, we
zoom on some individual peaks in the recurrence spectra, choosing
different sets of quantum defects but keeping the other parameters
(scaled energies, $\gamma$ range) constant, to observe the
presence of the on-node orbit and how its interplay with
core-scattering affects the amplitude of the recurrence peaks.

Fig.\ 3 displays the recurrence spectra for non-hydrogenic atoms
with $\mu_{l=1}=0.5$ [a]$,$ $\mu_{l=1}=0.25$ [b] and
$\mu_{l=1}=0.1$ [c], at $\epsilon=-0.3$, around the peaks labelled
$3$ and $1+3$ in Fig.\ 2. According to the standard treatment
(solid line), peak $3$ is produced by the $R_{2}^{1}$ ``pac-man'' orbit
(the shapes and characteristics of the orbits mentioned here are
given in Table I and Fig.\ 6 of \cite{mdm02}; $R_{2}^{1}$ has
bifurcated from $R_{2}$ at a slightly lower energy, and thus the
two orbits have nearly the same scaled action), and $1+3$ results
from the combination of the $V_{1}^{1}$ ``balloon'' orbit (peak
$1$) and $R_{2}^{1}$ through core-scattering. The mismatch for the
peak $3$ arises from interference between the contributions of
the $R_{2}^{1}$ and the
on-node $R_{2}$ orbit; indeed, including the on-node orbit in the
semiclassical calculations results in excellent agreement with the
quantum result. The peak $3$ thus results from the interference of
primitive orbits and accordingly does not depend on the value of
the quantum defect; however the peak $1+3$ does depend on the
quantum defect and vanishes in the limit $\mu_{l=1}\rightarrow0$;
the contribution of the on-node orbit in $1+3$ is seen to be
important (in absolute terms) only provided the quantum defect is
large. Note that in principle we should also have taken into
account the first and third repetitions of the perpendicular
orbit, but their corresponding amplitudes are very small, so these
orbits have a negligible contribution to the recurrence spectra.

The situation depicted in Fig.\ 4 is more involved: a close up of
the peak at $\tilde{S}=3.9$ (labelled $3+3$ in Fig.\ 2) is shown
for a molecule with quantum defects $\mu_{\Sigma}=0.5,$
$\mu_{\Pi}=0$ [a] and $\mu_{\Sigma }=0.5,$ $\mu_{\Pi}=0.5$ [b];
the peak $3+3$ arises from recurrences produced by different
orbits: the second return of $R_{2}^{1}$ and the fourth return
$R_{4}$ of the on-node perpendicular orbit, the combinations $R_{2}^{1}%
+R_{2}^{1}$, $R_{2}^{1}+R_{2}$ and $R_{2}+R_{2}$ via
core-scattering. The resulting peak amplitude depends both on the
quantum defects (which rule the core-scattering amplitudes) and on
the inclusion of the two on-node orbits: in the first case the
standard semiclassical result \emph{underestimates\/} the exact
quantum calculation, whereas in Fig.\ 4 (b) the standard
semiclassical result \emph{overestimates\/} the correct recurrence
strength. Adding the contribution of the on-node orbits in the
semiclassical treatment results in both cases in a better
agreement with the quantum calculations.

Fig.\ 5 displays the peak labelled $\mathrm{a}+3$ in Fig.\ 2 but for the choice
of quantum defects $\mu_{\Sigma}=0.5,$ $\mu_{\Pi}=0$.  This peak results
from the inelastic scattering between $R_{2}^{1}$ at $\epsilon=-0.3$ and the
perpendicular orbit associated with the core state $N=2,$ $m=\pm1$ at
$\epsilon=-0.8.$ Again, the standard closed-orbit result underestimates the
recurrence strength and the inclusion of the first-order suppressed on-node
orbit improves the agreement with the quantum results.

\begin{figure}[tb]
\includegraphics[height=2.8in,width=1.85in]{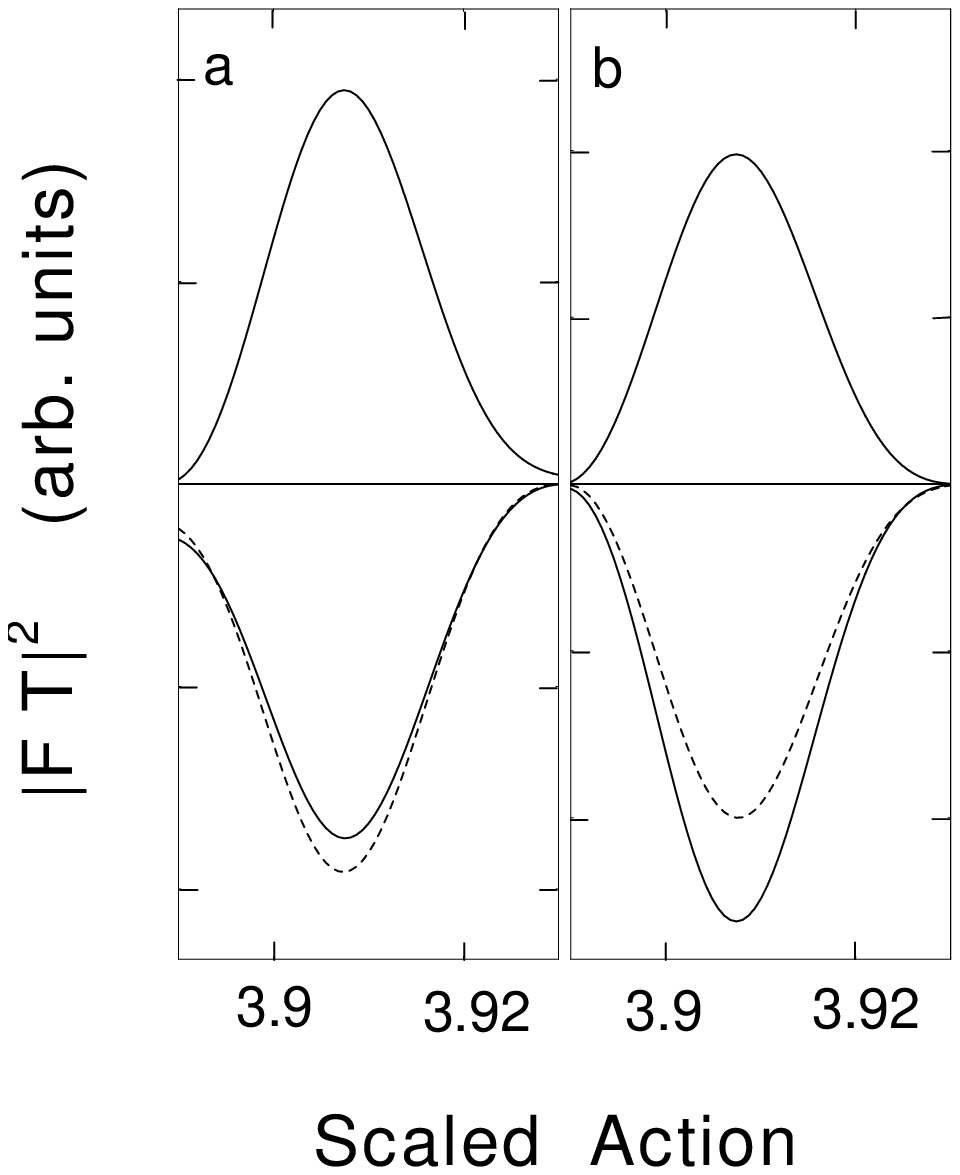}
\caption[]{The peak labelled $3+3$ in Fig.\ 2 is shown for a
molecule with quantum defects $\mu_{\Sigma}=0.5,$ $\mu_{\Pi}=0$
[a] and $\mu_{\Sigma }=0.5,$ $\mu_{\Pi}=0.5$ [b]. The inclusion of
the higher order contributions (broken line) gives a better
agreement with the quantum calculations (top) than the standard
semiclassical formalism (solid line upside-down). \label{fig.4}}
\end{figure}

Finally, Fig.\ 6 shows a portion of the recurrence spectrum for the
hydrogen atom at $\epsilon=-0.55,$ $M=1,$ in the range
$\gamma^{-1/3}=[30,240]$. We have zoomed the peaks at
$\tilde{S}=0.95$ and $\tilde{S}=1.91$ which are due to the first
and second returns of the classically ``non-existing'' forbidden
parallel orbit. Note that the peak at $\tilde{S}=0.95$ sits on the
right shoulder of the much stronger $R_{1}$ orbit, whereas the
second return at $\tilde {S}=1.91$ is sufficiently isolated. The
quantum calculation for $M=1$ thus displays peaks for orbits which
classically ``do not exist'', at the actions of the corresponding
$M=0$ parallel orbit. The standard semiclassical treatment (solid
line) can not obviously account for those peaks, but including Eq.
(\ref{31}), which takes into account higher order contributions,
yields an excellent agreement with the quantum results, since
those forbidden orbits contribute, albeit modestly, to the
photoabsorption spectrum.
\begin{figure}[tb]
\includegraphics[height=2.8in,width=1.25in]{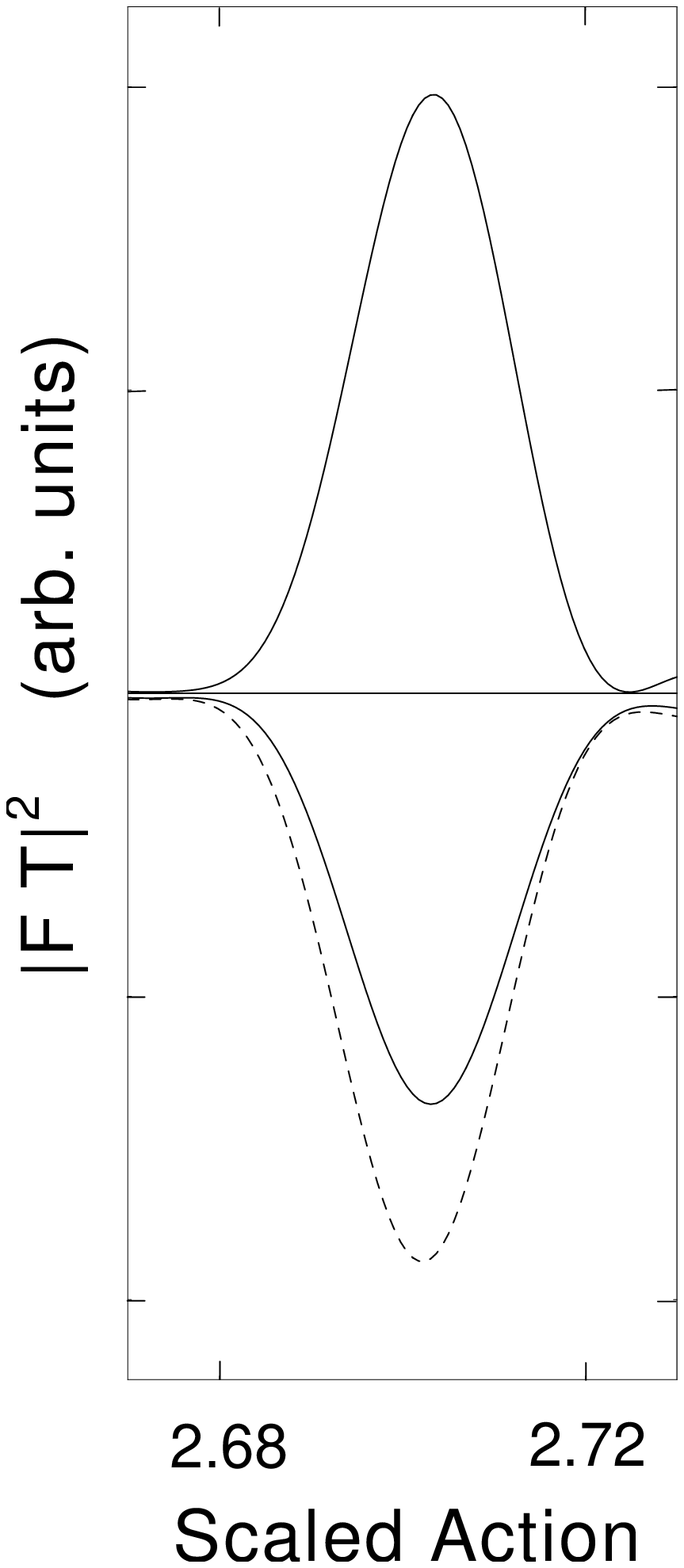}
\caption[]{The peak labelled $a+3$, due to inelastic
core-scattering in Fig.\ 2 is shown for the choice of quantum
defects $\mu_{\Sigma}=0.5,$ $\mu_{\Pi}=0$. The inclusion of the
higher order contributions (broken line) gives a better agreement
with the quantum calculations (top) than the standard
semiclassical formalism (solid line upside-down). \label{fig.5}}
\end{figure}

\section{Discussion and Conclusion}

The feature developed in this paper is one of the many refinements
that can be undertaken to improve a semiclassical formalism such
as closed orbit theory. Some processes are forbidden on purely
classical grounds (e.g., the above-barrier reflection of excited
lithium in an electric field which results in very broad
resonances in the absorption spectrum \cite{delos etal01}) whereas
other processes are semiclassically distorted (e.g., diverging
amplitudes at bifurcations).

\begin{figure}[tb]
\includegraphics[height=2.5in,width=3.2in]{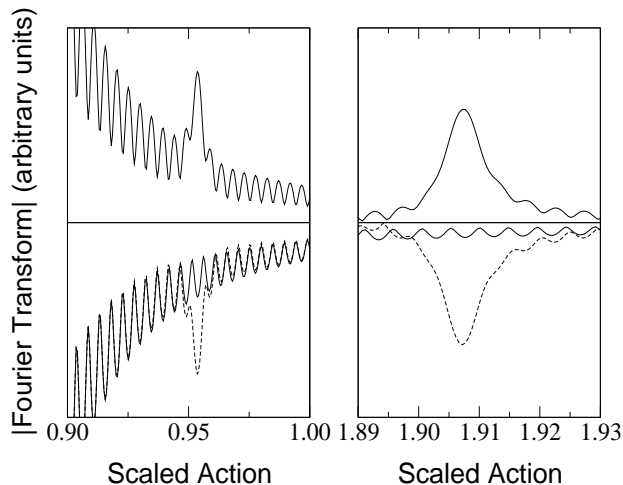}
\caption[]{Recurrence spectrum for the hydrogen atom at
$\epsilon=-0.55,$ $M=1,$ in the range $\gamma^{-1/3}=[30,240]$.
The plot focuses on the first and second repetitions of the
first-order suppressed ``parallel" orbit, which is clearly visible
on the quantum calculations (top). This feature is absent from the
standard semiclassical calculations (bottom, solid line), but the
inclusion of the higher order contributions (bottom, broken line)
results in an excellent agreement with the quantum calculations.
Note that the peak at $\tilde{S}=0.95$ sits on the right shoulder
of the much stronger $R_{1}$ orbit; the oscillations are due to
the finite range of the Welch-windowed Fourier transform.
\label{fig.6}}
\end{figure}
The r\^{o}le of the perpendicular on-node orbit was first observed
in calculations for the diamagnetic hydrogen atom at low scaled
energy ($\epsilon\simeq-0.7$) \cite{shaw etal95}. Subsequent
high-resolution experiments on helium in a magnetic field in the
same dynamical regime did not clearly detect the on-node orbits
(they were within the experimental noise) \cite{karremans etal99}.
We have given a simpler derivation of the contribution of these
first-order suppressed orbits, and our numerical results indicate
that on-node orbits are more likely to be detected in
non-hydrogenic atomic or molecular systems with strong quantum
defects. At low scaled energies, peaks resulting from the
core-scattering of the on-node orbit with a strong allowed orbit
could be more easily detected; at higher scaled energies, the
on-node orbit is most likely to affect the amplitude of peaks due
to typical allowed orbits.

The presence of contributions in the quantum photoabsorption
spectra which were not correlated with any classical orbit was
observed in calculations for non-hydrogenic atoms with $m\neq0$ in
an electric field by Robicheaux and Shaw \cite{robicheaux shaw98};
these contributions were coined ``recurrences without closed
orbits'' because they appear at the scaled action of the parallel
orbit which only exists classically when $L_{z}=0$ and should
therefore be absent in an $m\neq0$ recurrence spectrum. These
authors also gave an ad-hoc semiclassical formula akin to the
on-node correction which resulted in a poor agreement with the
quantum calculations.\ Main \cite{main99} later pointed out that,
for small but nonvanishing $L_{z}$, periodic orbits having nearly
the same action as the $L_{z}=0$ parallel orbit do exist; it
was unclear however whether the ``recurrences without closed
orbits'' could be attributed to such orbits, in particular because
the starting point of these orbits is several atomic units away
from the core. Our formula Eq. (\ref{31}) correctly accounts for
the peaks in the recurrence spectrum associated with these apparently
non-existing orbits; the $\hbar$ dependence is different to that of
the suppressed on-node orbits. The physical picture is similar
in both cases: just as Eq. (\ref{29}) accounts for close neighbors
to the on-node orbit, which are not closed at the origin but carry
a portion of the wavefunction back to the core region, Eq.
(\ref{31}) takes into account non-radial orbits close to the $z$
axis which also give rise to recurrences by carrying the
wavefunction from and into the core region.

To conclude, we have seen that first-order forbidden processes can be included
within Closed-orbit theory in a simple and unified manner by elementary
manipulations of the stationary phase integral, which yield a higher-order
$\hbar$ dependence. In passing, we have shown that the zero degree orbit,
which has always required special treatment, is in fact a case calling for
a refined stationary phase integration. Analogous manipulations of the
stationary phase integral of the Green's function were performed in
\cite{granger greene00} to obtain an improved semiclassical long-range
scattering matrix for Rydberg atoms in fields. Our method provides a
convenient and effective way of including non-radial and non-closed
trajectories that nevertheless contribute to the photoabsorption
spectra of Rydberg atoms and molecules in fields without the need to
to calculate explicitly the involved classical dynamics of those
trajectories. The validity of the method was assessed by comparing our
semiclassical results to quantum calculations for Rydberg atoms and
molecules in an external magnetic field.

\appendix
\section{}
We briefly work out the integral needed to determine the contribution of an
orbit lying on the node of a wavefunction in Sec. III,
\begin{equation}
\int_{0}^{\pi}d\bar{\theta}_{f}(\bar{\theta}_{f}-\theta_{fk})^{2}\exp
(i\sqrt{r_{f}/8}(\bar{\theta}_{f}-\theta_{fk})^{2}/\hbar).\label{a1}%
\end{equation}
This integral can be integrated directly but, for present purposes, it is
convenient to express it in terms of sine and cosine Fresnel integrals and
take the limit for the range in which the standard stationary phase
approximation holds for the usual orbits. For example, in the neighborhood of
$\theta_{fk}$, the real part of (\ref{a1}) can be expressed in the form%
\begin{align}
I(n)& =2\int_{\theta _{fk}}^{\theta _{fk}+\varepsilon _{n}}d\bar{\theta}_{f}(%
\bar{\theta}_{f}-\theta _{fk})^{2}\cos (\sqrt{r_{f}/8}(\bar{\theta}%
_{f}-\theta _{fk})^{2}/\hbar )  \label{a3} \\
& =\left[ -2^{3/4}\hbar ^{3/2}r^{-3/4}\pi ^{1/2}\right] \left\{ 2\mathcal{S}(%
\sqrt{1+2n})-2\sqrt{1+2n}\cos \pi n\right\} ,  \label{a4}
\end{align}%
where $\mathcal{S}$ is the sine Fresnel integral and
$\varepsilon _{n}=2^{1/4}r^{-1/4}\pi ^{1/2}\hbar ^{1/2}\sqrt{1+2n}$
with $n$
a real number $n>-1/2$. For large half integer values of $n,$ $%
I(n+1)-I(n)\approx 0$ and $\mathcal{S}(\sqrt{1+2n})\sim 1/2$.
$I(n)$ can then be approximated by the term between square
brackets in Eq. (\ref{a4}).\
This is consistent with having neglected terms of order $(\bar{\theta}%
_{f}-\theta _{fk})^{4}$ in Eq. (\ref{a1}) provided $\hbar
\rightarrow 0.$ The imaginary part of Eq. (\ref{a1}) is treated in
the same way by writing
the result in terms of the cosine Fresnel integral $\mathcal{C}(x)$. Hence%
\begin{equation}
\int_{0}^{\pi }d\bar{\theta}_{f}(\bar{\theta}_{f}-\theta _{fk})^{2}\exp (i%
\sqrt{r_{f}/8}(\bar{\theta}_{f}-\theta _{fk})^{2}/\hbar )\approx
\frac{\hbar ^{3/2}\pi ^{1/2}e^{3i\pi /4}}{r^{3/4}2^{-5/4}}.
\label{a6}
\end{equation}
Note that this result is independent of the value of $\theta_{fk}$,
provided $\theta _{fk}\neq0$.

\section{}
A closed form expression for the integrals%
\begin{equation}
I_{1}(\varepsilon )=\int_{0}^{\varepsilon }d\bar{\theta}_{f}\sin \bar{\theta}%
_{f}\exp (is\bar{\theta}_{f}^{2})  \label{b1}
\end{equation}%
and%
\begin{equation}
I_{2}(\varepsilon )=\int_{0}^{\varepsilon }d\bar{\theta}_{f}\sin \bar{\theta}%
_{f}\bar{\theta}_{f}^{2}\exp (is\bar{\theta}_{f}^{2}),  \label{b2}
\end{equation}%
with $s$ real, are obtained in the limit in which the standard
stationary phase approximation holds for the usual orbits by
replacing the upper bound by $\varepsilon \rightarrow \infty $.
Then Eqs. (\ref{b1}) and (\ref{b2}) are given in terms of infinite
series \cite{integral table1}, which are actually representations
of special functions.\ Choosing for simplicity a representation in
terms of Fresnel
integrals, Eq. (\ref{b1}) becomes in this limit%
\begin{equation}
I_{1}(\infty )=\left( \frac{\pi }{2s}\right) ^{1/2}\exp i\left( -\frac{1}{4s}%
+\frac{\pi }{2}\right) \left\{ \mathcal{C}\left[ (2\pi s)^{-1/2}\right] +i%
\mathcal{S}\left[ (2\pi s)^{-1/2}\right] \right\} ,
\end{equation}%
whereas for Eq. (\ref{b2}) we have%
\begin{equation}
I_{2}(\infty )=-\frac{1}{4s^{2}}\left\{ 1+\left( \frac{\pi
}{2s}\right)
^{1/2}\left( 2s-i\right) \exp \left( -\frac{i}{4s}\right) \left[ \mathcal{C}%
\left[ (2\pi s)^{-1/2}\right] +i\mathcal{S}\left[ (2\pi s)^{-1/2}\right] %
\right] \right\} .
\end{equation}%
When $s\rightarrow \infty ,$ to first order only
$\mathcal{C}\left[ (2\pi s)^{-1/2}\right] \sim (2\pi s)^{-1/2}$
contributes to $I_{1}(\infty )$ whereas for $I_{2}(\infty )$ the
term between the braces simply gives 2.


\begin{thebibliography}{99}

\bibitem{du delos88}
M.\ L.\ Du and J.\ B.\ Delos, Phys.\ Rev.\ A {\bf 38}, 1913 (1988).

\bibitem{dando etal95} P.\ A.\ Dando, T.\ S.\ Monteiro, D.\ Delande,
and K.\ T.\ Taylor, Phys.\ Rev.\ A {\bf 54}, 127 (1996).

\bibitem{m&m01} A.\ Matzkin and T.\ S.\ Monteiro,
Phys.\ Rev.\ Lett.\ \textbf{87}, 143002 (2001).

\bibitem{mdm02}
A.\ Matzkin, P.\ A.\ Dando, and T.\ S.\ Monteiro, Phys.\ Rev.\ A \textbf{66},
0134XX (2002).

\bibitem{sadovskii delos96}
D.\ A.\ Sadovskii and J.\ B.\ Delos, Phys.\ Rev.\ E \textbf{54}, 2033
(1996).

\bibitem{main wunner97}
J.\ Main and G.\ Wunner, Phys.\ Rev.\ A \textbf{55}, 1743 (1997).

\bibitem{shaw etal95} J.\ A.\ Shaw, J.\ B.\ Delos,
M.\ Courtney, and D.\ Kleppner, Phys.\ Rev.\ A \textbf{52}, 3695 (1995).

\bibitem{robicheaux shaw98}%
F.\ Robicheaux and J.\ Shaw, Phys.\ Rev.\ A \textbf{58}, 1043 (1998)

\bibitem{gao delos92}%
J.\ Gao and J.\ B.\ Delos, Phys.\ Rev.\ A \textbf{46}, 1455 (1992).

\bibitem{friedrich wintgen89} H.\ Friedrich and D.\
Wintgen, Phys.\ Rep.\ \textbf{183} , 37 (1989).

\bibitem{hupper etal96}
B.\ H\"{u}pper, J.\ Main and G.\ Wunner, Phys.\ Rev.\ A
\textbf{53}, 744 (1996).

\bibitem{schweizer etal93}
W.\ Schweizer, R.\ Niemeier, G.\ Wunner and H.\ Ruder, Z.\ Phys.\ D
\textbf{25}, 95 (1993).

\bibitem{delos etal01}
J.\ B.\ Delos, V.\ Kondratovich, D.\ M.\ Wang, D.\ Kleppner, and
N.\ Spellmeyer, Phys. Scripta T \textbf{90} (2001).

\bibitem{karremans etal99}
K.\ Karremans, W.\ Vassen, and W.\ Hogervorst, Phys.\ Rev.\ A
\textbf{60}, 2275 (1999).

\bibitem{main99}%
J.\ Main, Phys.\ Rev.\ A \textbf{60}, 1726 (1999).

\bibitem{granger greene00}%
B.\ E.\ Granger and C.\ H.\ Greene, Phys.\ Rev.\ A \textbf{62},
012511 (2000).

\bibitem{integral table1} I.\ S.\ Gradshteyn and I.\ M.\ Ryzhik,
\emph{Table of Integrals, series and products\/} (Academic Press,
Boston, 1994), Secs. 3.8-3.9.

\end{thebibliography}
\end{document}